\documentclass[aps,prl,twocolumn,showpacs,showkeys,superscriptaddress]{revtex4-1}

\usepackage[pdftex]{graphicx}
\usepackage{natbib}
\usepackage{amsmath}
\usepackage{color}

\newcommand{\one}{Fig.~\ref{f1}}
\newcommand{\two}{Fig.~\ref{f2}}
\newcommand{\three}{Fig.~\ref{f3}}
\newcommand{\sto}{SrTiO$_3$} 
\newcommand{\pcm}{cm$^{-2}$}
\newcommand{\ep}{\textit{e-p}}

\begin{document}

\title{Tailoring the nature and strength of electron-phonon interactions in the SrTiO$_3$(001) two-dimensional electron liquid}

\author{Z. Wang}
\affiliation{Swiss Light Source, Paul Scherrer Institut, CH-5232 Villigen PSI, Switzerland}
\affiliation{Department of Quantum Matter Physics, University of Geneva, 24 Quai Ernest-Ansermet, 1211 Geneva 4, Switzerland}
\author{S. McKeown Walker}
\affiliation{Department of Quantum Matter Physics, University of Geneva, 24 Quai Ernest-Ansermet, 1211 Geneva 4, Switzerland}
\author{A. Tamai}
\affiliation{Department of Quantum Matter Physics, University of Geneva, 24 Quai Ernest-Ansermet, 1211 Geneva 4, Switzerland}
\author{Y. Wang}
\affiliation{Stanford Institute for Materials and Energy Sciences, SLAC National Accelerator Laboratory, Menlo Park, California 94025, USA}
\affiliation{Department of Applied Physics, Stanford University, Stanford, CA 94305, USA}
\author{Z. Ristic}
\affiliation{Swiss Light Source, Paul Scherrer Institut, CH-5232 Villigen PSI, Switzerland}
\author{F.Y. Bruno}
\affiliation{Department of Quantum Matter Physics, University of Geneva, 24 Quai Ernest-Ansermet, 1211 Geneva 4, Switzerland}
\author{A. de la Torre}
\affiliation{Department of Quantum Matter Physics, University of Geneva, 24 Quai Ernest-Ansermet, 1211 Geneva 4, Switzerland}
\author{S. Ricc\`o}
\affiliation{Department of Quantum Matter Physics, University of Geneva, 24 Quai Ernest-Ansermet, 1211 Geneva 4, Switzerland}
\author{N.C. Plumb}
\affiliation{Swiss Light Source, Paul Scherrer Institut, CH-5232 Villigen PSI, Switzerland}
\author{M. Shi}
\affiliation{Swiss Light Source, Paul Scherrer Institut, CH-5232 Villigen PSI, Switzerland}
\author{P. Hlawenka}
\affiliation{Helmholtz-Zentrum Berlin f\"ur Materialien und Energie GmbH}
\author{J. S\'anchez-Barriga}
\affiliation{Helmholtz-Zentrum Berlin f\"ur Materialien und Energie GmbH}
\author{A. Varykhalov}
\affiliation{Helmholtz-Zentrum Berlin f\"ur Materialien und Energie GmbH}
\author{T.K. Kim}
\affiliation{Diamond Light Source, Harwell Campus, Didcot, United Kingdom}
\author{M. Hoesch}
\affiliation{Diamond Light Source, Harwell Campus, Didcot, United Kingdom}
\author{P.D.C. King}
\affiliation{SUPA, School of Physics and Astronomy, University of St Andrews, St Andrews, Fife KY16 9SS, United Kingdom}
\author{W. Meevasana}
\affiliation{School of Physics, Suranaree University of Technology, Nakhon Ratchasima, 30000, Thailand}
\author{U. Diebold}
\affiliation{Institute of Applied Physics, Vienna University of Technology, Wiedner Hauptstrasse 8-10/134, A-1040 Vienna, Austria}
\author{J. Mesot}
\affiliation{Swiss Light Source, Paul Scherrer Institut, CH-5232 Villigen PSI, Switzerland}
\affiliation{Institute of Condensed Matter Physics, \'Ecole Polytechnique F\'ed\'erale de Lausanne (EPFL), CH-1015 Lausanne, Switzerland}
\affiliation{Laboratory for Solid State Physics, ETH Z\"urich, CH-8093 ZŸrich, Switzerland}
\author{B. Moritz}
\affiliation{Stanford Institute for Materials and Energy Sciences, SLAC National Accelerator Laboratory, Menlo Park, California 94025, USA}
\author{T.P. Devereaux}
\affiliation{Stanford Institute for Materials and Energy Sciences, SLAC National Accelerator Laboratory, Menlo Park, California 94025, USA}
\affiliation{Geballe Laboratory for Advanced Materials, Stanford University, Stanford, CA 94305, USA}
\author{M. Radovic}
\affiliation{Swiss Light Source, Paul Scherrer Institut, CH-5232 Villigen PSI, Switzerland}
\affiliation{SwissFEL, Paul Scherrer Institut, CH-5232 Villigen PSI, Switzerland}
\author{F. Baumberger}
\affiliation{Department of Quantum Matter Physics, University of Geneva, 24 Quai Ernest-Ansermet, 1211 Geneva 4, Switzerland}
\affiliation{Swiss Light Source, Paul Scherrer Institut, CH-5232 Villigen PSI, Switzerland}
\affiliation{SUPA, School of Physics and Astronomy, University of St Andrews, St Andrews, Fife KY16 9SS, United Kingdom}





\maketitle


\textbf{Surfaces and interfaces offer new possibilities for tailoring the many-body interactions that dominate the electrical and thermal properties of transition metal oxides~\cite{Mannhart:sc10,Lee:nat14,Zubko:2011arcmp}. 
Here, we use the prototypical two-dimensional electron liquid (2DEL) at the SrTiO$_3$(001) surface~\cite{Ohtomo:nat04,Meevasana:natm11,Santander:nat11,King:natc2014} to reveal a remarkably complex 
evolution of electron-phonon coupling with the tunable carrier density of this system.
At low density, where superconductivity is found in the analogous 2DEL at the LaAlO$_3$/SrTiO$_3$ interface~\cite{Thiel:sc06, Reyren:sc07, Caviglia:nat08,Richter:nat13,Cheng:nat15,Boschker:srep15},
our angle-resolved photoemission data show replica bands separated by 100~meV from the main bands. This is a hallmark of a coherent polaronic liquid and implies long-range coupling to a single longitudinal optical phonon branch. In the overdoped regime the preferential coupling to this branch decreases and the 2DEL undergoes a crossover to a more conventional metallic state with weaker short-range electron-phonon interaction. These results place constraints on the theoretical description of superconductivity and allow for a unified understanding of the transport properties in \sto-based 2DELs.
}

Carrier concentration is a key parameter defining the ground state of correlated electron systems. At the LaAlO$_3$/SrTiO$_3$ interface, the 2DEL density can be tailored by field-effect gating. 
As the system is depleted of carriers, its ground state evolves from a high mobility 2DEL~\cite{Ohtomo:nat04}, into a two-dimensional superconductor~\cite{Thiel:sc06, Reyren:sc07, Caviglia:nat08} with pseudogap behaviour~\cite{Richter:nat13} and possible pairing above T$_c$\cite{Cheng:nat15}.
An analogous 2DEL can be induced by doping the (001) surface of \sto. As for the interface, the surface 2DEL is confined by a band bending potential in \sto{} and consists of an orbitally polarized ladder of quantum confined Ti $t_{2g}$ electrons that are highly mobile in the surface plane~\cite{Meevasana:natm11,Santander:nat11,King:natc2014,Plumb:prl14}. Thus far, the surface 2DEL has only been studied at carrier densities around $2\times10^{14}$~\pcm, approximately a factor of five higher than typically observed at the LaAlO$_3$/SrTiO$_3$ interface~\cite{Meevasana:natm11,Santander:nat11,King:natc2014}.
In the following, we present new ARPES data extending to lower carrier densities that are directly comparable to the LaAlO$_3$/SrTiO$_3$ interface. We achieve this by preparing \sto{}(001) wafers \textit{in situ}, which results in well ordered clean surfaces that can be studied by ARPES over extended timescales as they are less susceptible to the UV induced formation of charged oxygen vacancies reported for cleaved \sto~\cite{Meevasana:natm11,King:natc2014,Wang:pnas14,McKeown:am15}. Details of the sample preparation are given in the methods section. 

\begin {figure*}[t]
\includegraphics [width=6.5 in,clip] {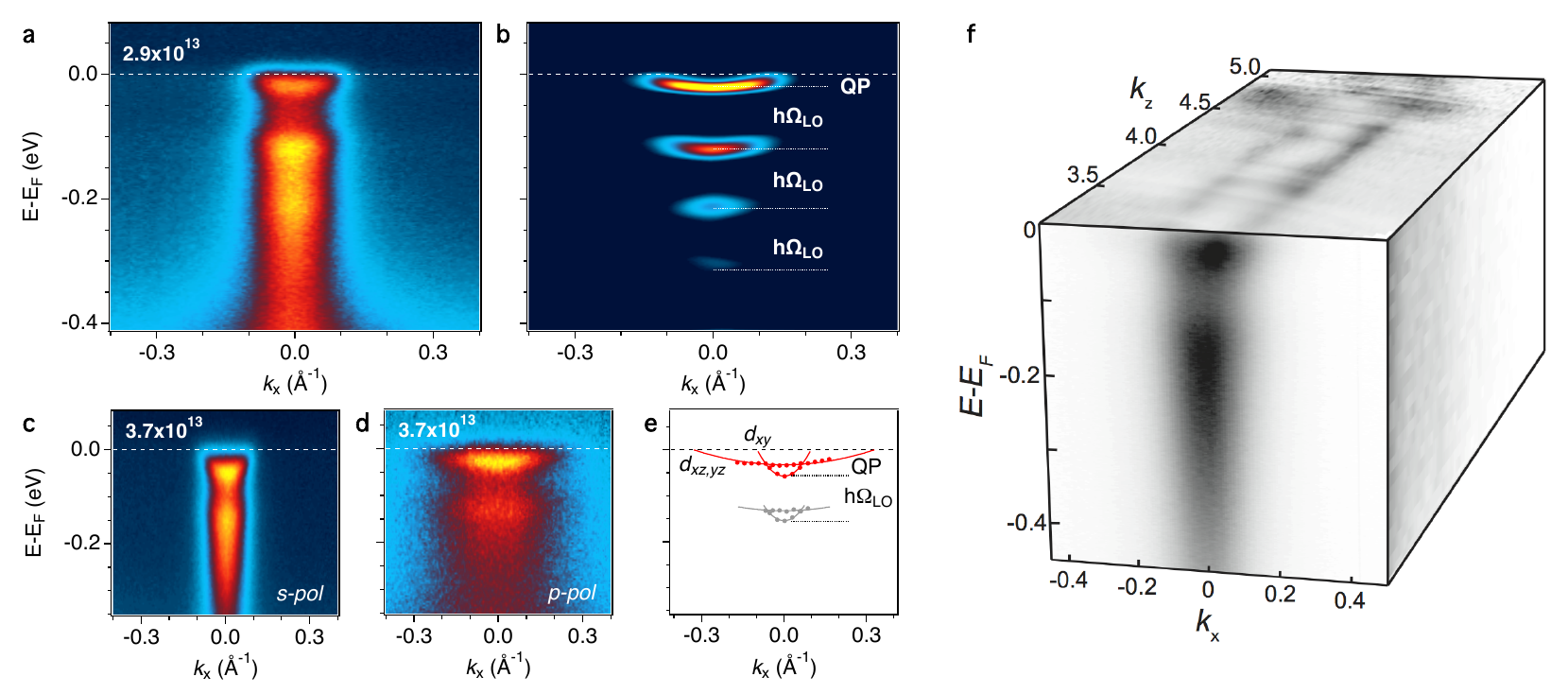}
 \caption{
\textbf{A two-dimensional liquid of large polarons in SrTiO$_3$.} \textbf{a,b} Energy-momentum intensity map and curvature plot for a 2DEL with $n_{2D} = 2.9\cdot10^{13}$~cm$^{-2}$ taken at a photon energy of 44~eV with $s$-polarization. Note the dispersive replica bands at higher binding energy arising from strong coupling to the LO$_4$ phonon branch of \sto. \textbf{c,d} Data taken on a sample with $n_{2D} = 3.7\cdot10^{13}$~cm$^{-2}$ using $s$ and $p$ polarized light with $h\nu=85$~eV to selectively excite the light $xy$ and heavy $xz/yz$ orbitals. \textbf{e} Dispersion of the main and first replica bands extracted from the data in \textbf{c,d}. \textbf{f} Photon energy dependent measurements showing the lack of dispersion along $k_z$.
All data were measured in the second Brillouin zone to avoid the minimum of the matrix elements at normal emission.
}
\label{f1}
\end{figure*}

\one~\textbf{a} shows an energy-momentum intensity map for a 2DEL with a carrier density of $n_{\rm{2D}}\approx$\,2.9\,$\cdot$10$^{13}$\,cm$^{-2}$ estimated from the Luttinger volume of the first light subband and the two equivalent heavy subbands (see Supplementary Information section 2).
The most striking feature of this data are replica bands at higher binding energy following the dispersion of the primary quasiparticle (QP) bands. The replica bands are all separated by $\approx100$~meV and progressively loose intensity but can be visualized up to the third replica in the curvature plot shown in \one~\textbf{b}. 
From the equal energy spacing of the replica bands we can rule out that they represent distinct subbands arising from quantum confinement~\cite{Meevasana:natm11,Santander:nat11,King:natc2014,McKeown:am15}.
We thus interpret the replicas as shake-off excitations involving a single non-dispersive bosonic mode coupling electrons of momentum $\mathbf{k}$ and $\mathbf{k}+\mathbf{q}$. From its energy of $\approx100$~meV, we can identify this mode as the highest frequency longitudinal optical phonon branch (LO$_4$) of \sto~\cite{Gervais:prb93}. 
A plasmon mode in the same energy range~\cite{Gervais:prb93,Chang:prb2010} can be excluded from the negligible density dependence of the mode frequency observed in \two{}.
Our ARPES data show that the coupling to this mode largely preserves the dispersion in the replica bands and thus must be restricted to small values of $\mathbf{q}$. This is a hallmark of Fr\"ohlich polarons, quasiparticles formed by an excess electron dressed by a polarization cloud extending over several lattice sites that follows the charge as it propagates through the crystal~\cite{Devreese:rpp2009,Alexandrov:iop2003,Lee:nat14,Moser:prl13}. 
Such a large polaron state preserves band-like transport but has an increased effective mass $m^{*}$. From our ARPES data we can directly quantify this effect. Using a parabolic fit of the band dispersion in \one~\textbf{a} we find $m_{xy}^{*}\approx 1.4$~m$_e$. Using a bare band mass of $m_0=0.6$~m$_e$~\cite{King:natc2014,Wang:pnas14} this corresponds to a mass enhancement $m_{xy}^{*}/m_0\approx 2.3$, indicative of an intermediate \ep{} coupling strength.

In \one~\textbf{c-f} we investigate a sample with slightly higher carrier density.
Using $s$ and $p$ polarized light, respectively, we selectively excite electrons from the light $xy$ band with $\approx 50$~meV occupied bandwidth and a shallower heavy band derived from \textit{xz/yz} orbitals. 
Both bands show a clear peak-dip-hump line shape with a dispersive replica band as summarized in \one~\textbf{e}. This implies that the entire orbitally polarized 2DEL including the heavy \textit{xz/yz} states, which are believed to be important for superconductivity~\cite{Reyren:sc07,Caviglia:nat08,Richter:nat13}, is a polaronic liquid at low carrier densities.
The lifting of the orbital degeneracy by approximately 20~meV can be attributed to quantum confinement in the band bending potential which increases the energy of out-of-plane orbitals \cite{Salluzzo:prl09,King:natc2014}. The reduced dimensionality arising from quantum confinement is visualized directly in \one~\textbf{f} where we show the absence of dispersion along the surface normal $k_z$, characteristic of two-dimensional electronic states.

\begin {figure*}[t]
\includegraphics [width=6.7 in,clip] {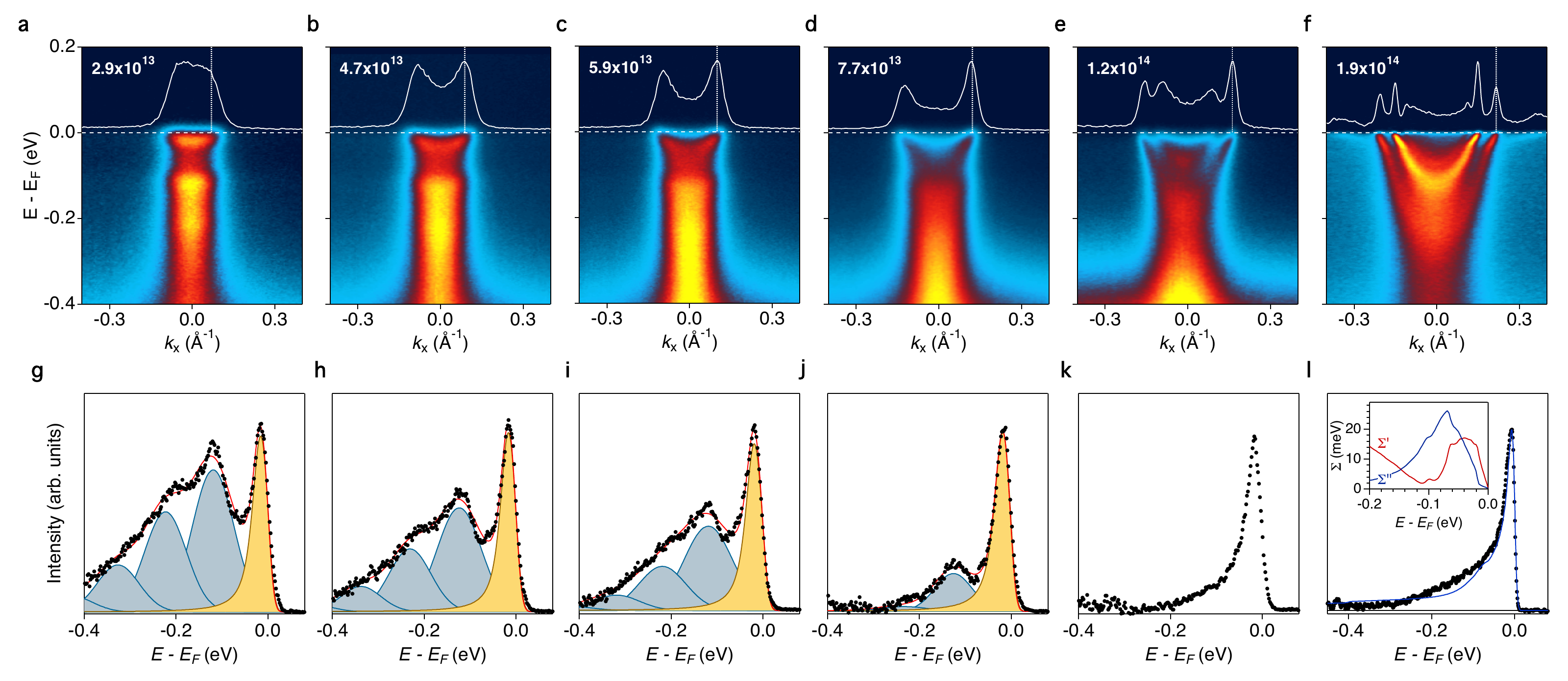}
 \caption{
\textbf{Evolution of the 2DEL spectral function with carrier concentration.} \textbf{a}-\textbf{f}, Raw energy-momentum intensity maps of 2DELs with increasing carrier concentration indicated in units of \pcm. \textbf{g}-\textbf{l}, Energy distribution curves at the Fermi wave vector indicated by a dashed white line in the corresponding image plots. An exponential background describing the tail of the in-gap state at $\approx-1.1$~eV has been subtracted from the raw EDCs (see Supplementary Information). In \textbf{g}-\textbf{j} we show fits to a Franck-Condon model with a single phonon mode. The coherent quasiparticle and incoherent phonon satellites are coloured in yellow and blue, respectively. \textbf{l} shows a calculation of the spectral function using the conventional Eliashberg \ep{} self-energy for the high-density limit found in Ref.~\cite{King:natc2014} and reproduced in the inset.}
\label{f2}
\end{figure*}

We now turn our attention to the strength of \textit{e-p} coupling. Its systematic evolution with carrier density in the \sto{}(001) 2DEL is shown in \two{}. Using $s$-polarized light, we resolve a single $d_{xy}$ QP band at low density. Energy distribution curves extracted at $k_{F}$ show a strongly reduced weight of the coherent QP and at least two phonon satellites each separated by $\approx100$~meV.
Empirically, we find that the experimental spectra are well described by a Franck-Condon model using a single phonon mode of $\approx 100$~meV.
Highly restricted fits using the characteristic Poisson distribution $I_n / I_{QP} = a_{c}^{2n}/n!$ for the intensity ratio of the $n$-th phonon satellite and the QP peak in the Franck-Condon model are shown in \two~\textbf{g-l}. Details of the fits are described in Supplementary Information. From this analysis we infer a quasiparticle residue $Z\approx0.2$ at the lowest density studied here. 
This is beyond the validity of the perturbation theory result for the Fr\"ohlich model of $Z=1-\alpha/2$~\cite{Lee:pr53,Mishchenko:prb00}, where $\alpha$ is the dimensionless coupling constant,
placing the SrTiO$_3$ 2DEL in a theoretically challenging regime of intermediate coupling. Moreover, the ratio of lattice energy to electron kinetic energy is neither small nor independent of doping, which excludes a Migdal-Eliashberg approach.
We therefore estimate $\alpha$ from the experimentally determined quasiparticle residue $Z$ using the results of diagrammatic quantum Monte Carlo simulations of the Fr\"ohlich model reported in Ref.~\cite{Mishchenko:prb00}. This gives $\alpha\approx2.8$ at the lowest carrier density, comparable to $\alpha=2-3$ reported for lightly doped bulk \sto{} based on an analysis of optical conductivity data~\cite{Mechelen:prl08,Devreese:prb10}.

As the carrier density increases, the Fermi wave vector increases monotonically, and new subbands become discernible. Concomitantly, the spectral weight of the phonon satellites weakens and for densities above $n_{2}\approx 9\cdot10^{13}$~\pcm{} they can no longer be resolved experimentally.
At high density \ep{} coupling is not only weaker but also of fundamentally different nature. This is illustrated in \two~\textbf{f,l}. In this regime, the quasiparticle dispersion shows a weak kink at an energy of $\approx 30$~meV and no signs of replica bands can be discerned in the raw data or in curvature plots (see Supplementary Information, Fig.~S4), providing direct evidence for a suppression of the long-range Fr\"ohlich interaction. 
Instead, consistent with Ref.~\cite{King:natc2014}, we find that the spectral function and \ep{} self-energy of the high-density 2DEL can be described by Migdal-Eliashberg theory with $\lambda\approx0.7$ and the same coupling to the entire phonon density of states.
 
\begin {figure}[t]
\includegraphics [width=3.4 in,clip] {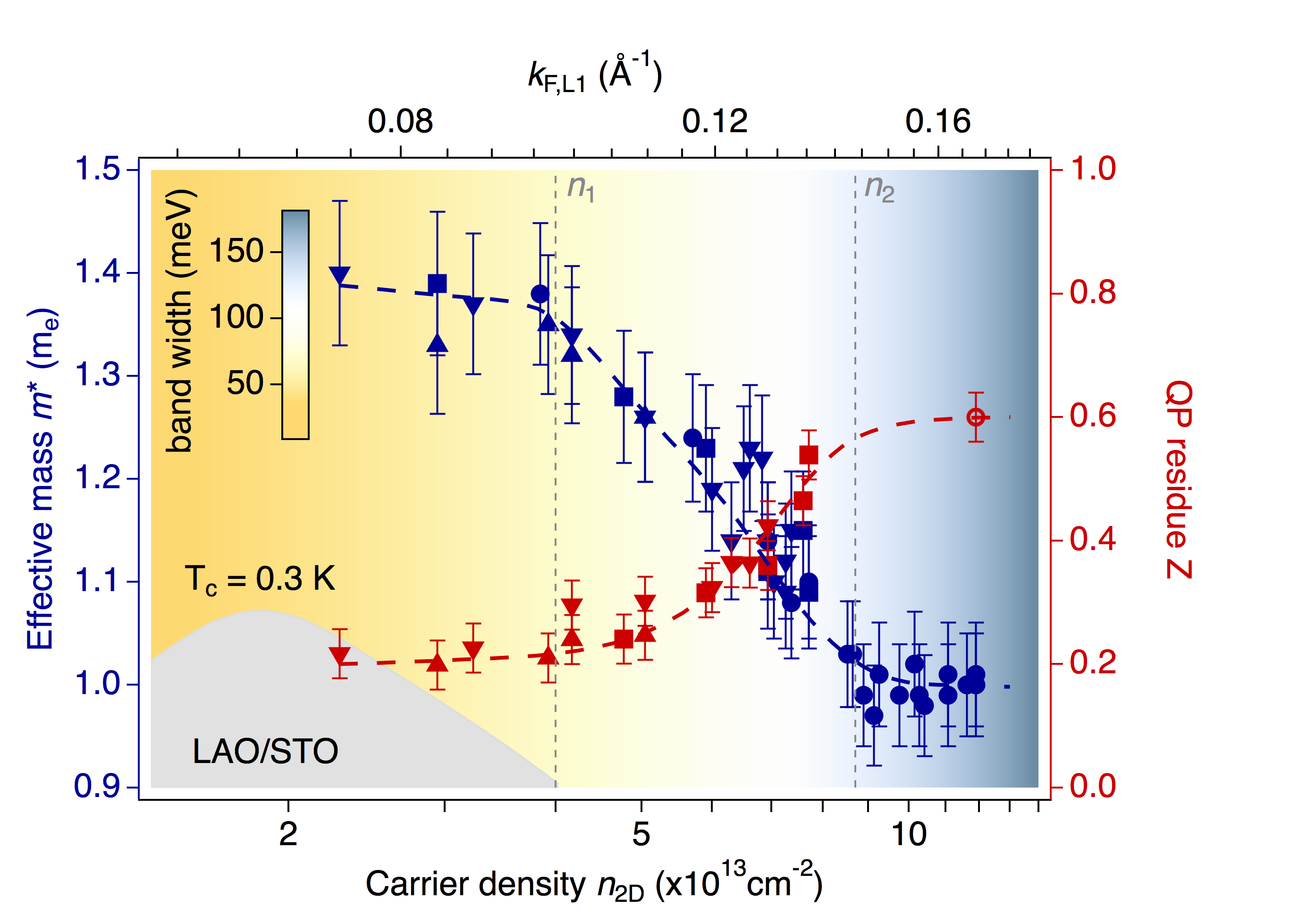}
 \caption{
\textbf{Effective mass and quasiparticle residue in the \sto{} 2DEL.} Evolution of the effective mass $m^{*}$ (blue symbols) and quasiparticle residue $Z$ (red symbols) with carrier density. Different symbols indicate data taken on substrates annealed at different temperature. Closed red symbols are obtained from Franck-Condon fits, while the last value with open symbol in the adiabatic Migdal-Eliashberg regime has been calculated from $Z=m_0 / m^{*}$. Error bars indicate the reproducibility of our results. An additional systematic error cannot be excluded.
The background color encodes the bare band width of the first light subband calculated from the experimentally determined $k_F$ shown in the top-axis, assuming a bare mass of $m_0 = 0.6$~m$_e$. Dashed lines are guides to the eye. The dome shaped superconducting phase observed at the LaAlO$_3$/SrTiO$_3$ interface is indicated in grey.
}
\label{f3}
\end{figure}

In order to track the crossover from the polaronic state to short-range \ep{} coupling, we analyse the quasiparticle residue and effective mass of the dispersive bands close to the Fermi level as a function of 2D carrier concentration. The results are summarized in \three{} where we plot $m^{*}$ obtained from fits to the QP dispersion and $Z(k_F)$ from fits to a Franck-Condon model as described above.
Three distinct regimes can be identified. Below $n_{1}\approx 4\cdot10^{13}$~\pcm{} both quantities depend weakly on carrier density and appear to saturate around $m^{*}\approx1.4$~m$_e$ and $Z(k_F)\approx0.2$, respectively. 
At intermediate carrier concentrations of $n_{2D} = 4 - 9\cdot10^{13}$~\pcm, the polaronic state persists, as is seen most directly from the spectra in \two~\textbf{h-j} showing a clear phonon satellite over this entire regime. 
However, approaching $n_{2}\approx 9\cdot10^{13}$~\pcm, where the phonon satellites are no longer resolved experimentally, the quasiparticle residue increases by more than a factor of two to $Z\approx 0.5$ and the effective coupling strength decreases to $\alpha\approx1.3$.
This is opposite to the trend expected for short-range \textit{e-p} coupling~\cite{Mishchenko:prl14} and thus strongly supports our identification of long-range interactions as described by the Fr\"ohlich model.
We note that the weak coupling to the LO$_4$ branch close to $n_2$ reported in \three{} has recently been confirmed for the LaAlO$_3$/SrTiO$_3$ interface 2DEL in a soft x-ray ARPES study reporting $Z\approx0.4$ for a sample with $n_{2D}\approx 8\cdot10^{13}$~\pcm, in excellent agreement with our findings~\cite{Cancellieri:arxiv15}.

The effective mass decreases more slowly than the quasiparticle residue and saturates at $m^{*}\approx1.0$~m$_e$ at high density. This slow decrease is characteristic of the Fr\"ohlich model where $Z < m_0 / m^{*}$.
For weak to intermediate coupling, the effective mass of Froehlich polarons can be approximated as $m^{*}/m_0 = 1/(1-\alpha/6)$~\cite{Devreese:rpp2009,Mishchenko:prb00}.
As shown in Supplementary Information Fig.~S6, this relation systematically underestimates the effective masses obtained directly from the quasiparticle dispersion but reproduces their trend as a function of density. 
We tentatively assign this behavior to the effect of electron-electron interactions, which is not fully included in the analysis of $Z$ in the polaronic regime.
Indeed, the factor between the two effective masses is approximately 1.3, which is consistent with the mass-enhancement due to electronic correlations estimated in Ref.~\cite{King:natc2014}. 
For densities above $n_{2}$, described by Migdal-Eliashberg theory, $Z=m/m^{*}$ suggesting that the quasiparticle residue saturates around $Z=0.6$ in the high-density 2DEL.


We attribute the breakdown of the polaronic state at high carrier density to improved electronic screening suppressing the long range Fr\"ohlich interaction. In order to estimate the cross-over from dielectric screening in the polaronic regime to predominantly electronic screening, we compare the polaron radius $r_p=(\hbar/2m^{*}\Omega_{\rm{LO,4}})^{-1/2}$~\cite{Devreese:rpp2009} with the electronic screening length. Using the experimentally determined parameters we find $r_p\approx6$~\AA{}. Given the relatively large dielectric constant of doped SrTiO$_3$, electronic screening is in the Thomas-Fermi regime and the screening length can be estimated from $r_{\rm{TF}}=(\epsilon\epsilon_{0}E_F/2e^{2}n_{3D})^{1/2}$ for a 3D electron liquid.
Assuming a uniform charge distribution over a 2DEL thickness of 3 unit cells and a static dielectric constant $\epsilon_0=100$, which provides a good description of the band-bending potential in the doped surface region~\cite{King:natc2014,McKeown:am15}, we estimate that $r_{\rm{TF}}=r_p$ for a density $n_{2D} \approx 5\cdot 10^{13}$~cm$^{-2}$, slightly above $n_1$ where $Z$ starts to increase progressively resulting in a reduced effective coupling constant $\alpha$. At higher carrier densities, electronic screening will rapidly become more important as $\epsilon_0$ decreases simultaneously with the increase in $n$. We note that the moderate electronic screening length $r_{\rm{TF}}$ justifies the above use of an expression for 3D electron liquids. This basic picture is confirmed by our calculations of the spectral function reported in supplementary Fig.~S5. Using exact diagonalization of a model Hamiltonian with constant coupling parameter, we find pronounced replica bands at low carrier density and a rapid suppression of the replicas as the carrier density is increased over the relevant range. These calculations thus reproduce the key experimental findings of our study and support the idea that increasing electronic screening drives the observed breakdown of the polaronic liquid at high carrier densities.

The above results demonstrate that \ep{} interaction in \sto{} based 2DELs is remarkably complex and strongly dependent on carrier density. This provides new insight into the superconducting pairing mechanism at the LaAlO$_3$/SrTiO$_3$ interface, which has so far eluded experimental investigation.
As shown in \three, superconductivity in LaAlO$_3$/SrTiO$_3$ interface 2DELs~\cite{Thiel:sc06, Reyren:sc07, Caviglia:nat08, Lin:prl14} occurs in the polaronic low-density regime and its suppression on the overdoped side coincides with a gradually decreasing coupling to the LO$_4$ branch. This supports the notion that superconductivity is phonon mediated~\cite{Klimin:prb14,Boschker:srep15,Gorkov:arxiv15} and suggests that the pairing potential is dominated by the exchange of high-frequency longitudinal phonons rather than soft modes.
We note that our experimental spectral functions do not exclude the formation of large bipolarons, which have been discussed early on in the context of superconductivity in polar oxides~\cite{Emin:prl89,Hohenadler:prb04}. 
Superconducting susceptibilities calculated within our exact diagonalization scheme reported in supplementary information section 5 provide additional insight. Although limitations in the Hilbert space size prohibit a quantitative comparison with experiment, the calculations clearly show that the dominant pairing channel has $s$-wave symmetry. Moreover, the saturation of the superconducting susceptibility with carrier density found in this model provides evidence for a competition between the opposite trends of density of states and effective \textit{e-p} coupling underlying dome-shaped superconductivity in SrTiO$_3$ 2DELs.
We note that coupling to the LO$_4$ branch of \sto{} with comparable strength was also invoked to explain the anomalously high superconducting critical temperature of FeSe monolayers on \sto{} substrates~\cite{Wang:cpl12,Lee:nat14}.


\textbf{Methods}
The Nb-doped (0.5\,wt$\%$) and La-doped (0.075\,wt$\%$) SrTiO$_3$(001) surfaces were prepared by mild Ar$^+$ sputtering (600~eV, 2~$\mu$A, 5~min) followed by annealing in $2\cdot10^{-6}$~mbar oxygen for 0.5~h at temperatures varying from 700 to 1000$^{\circ}$C as monitored by an infrared pyrometer. 
Consistent with measurements on \textit{in situ} prepared and cleaved \sto, we find no spectral weight in the entire bulk band gap on freshly prepared surfaces (see Supplementary Information, Fig.~1)~\cite{Wang:pnas14,Meevasana:natm11,McKeown:am15}. Quantum confined metallic states appear concomitant with a localized in-gap state at higher energy after exposure of the surface to synchrotron light~\cite{Meevasana:natm11,Wang:pnas14,McKeown:am15}. We find that the rate at which the 2DEL carrier density increases under the synchrotron beam increases with the annealing temperature and exploit this to stabilize low-carrier densities over the extended exposure times required for detailed ARPES measurements.
ARPES measurements were performed at the SIS beamline of the Swiss Light Source, the I05 beamline of Diamond Light Source and the $1^2$ beamline of BESSY II at the Helmholtz-Zentrum Berlin. Data was acquired at $T\approx20$~K with $h\nu=30-100$~eV and energy and angular resolutions of $10-25$~meV and $\approx0.2^{\circ}$, respectively.

\begin{acknowledgments}
\textbf{Acknowledgments}
We thank A. F\^ete, M. Grilli, L. Patthey, V. Strocov, J.-M. Triscone, D. van der Marel and Z. Zhong for discussions. This work was supported by the Swiss National Science Foundation (200021-146995). PDCK was supported by the UK-EPSRC (EP/I031014/1) and the Royal Society, UD by the ERC Advanced Grant 'OxideSurfaces' and WM by the Thailand Research Fund (TRF) under the TRF Senior Research Scholar, Grant No. RTA5680008.
We acknowledge Diamond Light Source for time on beamline I05 under proposal SI11741.
\end{acknowledgments}

\end{document}